\documentclass[twocolumn,showpacs,amsmath,amstex,amssymb,mathfonts,prl,superscriptaddress]{revtex4}
\usepackage{amsthm,amsfonts,graphicx,verbatim}
\usepackage{amsmath}
\usepackage{amssymb}
\usepackage{amsthm}
\usepackage{amsfonts}
\usepackage{listings}
\usepackage{enumerate}
\usepackage{latexsym}
\usepackage{psfrag}
\usepackage{bm}
\usepackage{graphicx}
\usepackage{subfigure}
\newcommand{\be}{\begin{equation}}
\newcommand{\ee}{\end{equation}}
\newcommand{\bea}{\begin{eqnarray}}
\newcommand{\eea}{\end{eqnarray}}

\newcommand{\la}{\langle}
\newcommand{\ra}{\rangle}

\renewcommand{\phi}{\varphi}
\renewcommand{\epsilon}{\varepsilon}

\begin{document}
\title{Interferometric approach to measuring band topology in 2D optical lattices}
\author{Dmitry A. Abanin} 
\affiliation{Physics Department, Harvard University, Cambridge,
Massachusetts 02138, USA}
\author{Takuya Kitagawa} 
\affiliation{Physics Department, Harvard University, Cambridge,
Massachusetts 02138, USA}

\author{Immanuel Bloch}
\affiliation{Max-Planck-Institut f\"ur Quantenoptik, 85748 Garching, Germany}
\affiliation{Ludwig-Maximilians-Universit\"at, 80799 M\"unchen, Germany}

\author{Eugene Demler} 
\affiliation{Physics Department, Harvard University, Cambridge,
Massachusetts 02138, USA}

\date{\today}
\begin{abstract}

Recently, optical lattices with non-zero Berry's phases of Bloch bands have been realized. New approaches for measuring Berry's phases and topological properties of bands with experimental tools appropriate for ultracold atoms need to be developed. In this paper, we propose an interferometric method for measuring Berry's phases of two dimensional Bloch bands. The key idea is to use a combination of Ramsey
interference and Bloch oscillations to measure Zak phases, i.e. Berry's phases for closed
trajectories corresponding to reciprocal lattice vectors. We demonstrate that this technique
can be used to measure Berry curvature of Bloch bands, the $\pi$ Berry's phase of Dirac points, and the first Chern number of topological bands. We discuss several experimentally feasible realizations of this technique, which make it robust against low-frequency magnetic noise. 

\end{abstract}

\pacs{67.85.-d, 03.65.Sq, 03.65.Vf }

\maketitle

{\bf Introduction.} Topology underlies many fundamental physical phenomena in two-dimensional materials, most notably the quantum and anomalous Hall effects~\cite{NiuReview,TKNN,Niu95,Haldane2004}. The topological features of a 2D Bloch band are determined by the Berry's (geometrical) phases~\cite{Berry} -- the phases picked up during an adiabatic motion of a particle along closed trajectories in quasi-momentum space. The energy bands are classified by an integer-valued topological invariant ${\cal{C}}$, the first Chern number, which is proportional to the Berry's phase for a trajectory enclosing a full Brillouin zone (BZ). A filled band with Chern number ${\cal{C}}$ is characterized by the quantized Hall conductitivy ${\cal{C}} e^2/h$~\cite{TKNN,NiuReview}. 

Over the past few years, the interest in topological properties of 2D systems was strongly revived, following the discovery of new materials, including graphene~\cite{Castro09} and topological insulators~\cite{Hasan}, where the Berry's phases play an important role in defining the transport properties. For example, the band structure of graphene hosts two massless Dirac points at the corners of the BZ. A trajectory enclosing a Dirac point is characterized by the Berry's phase $\pi$, and it is this $\pi$ Berry's phase that lies at the heart of new phenomena observed in graphene, such as the half-integer quantum Hall effect and weak anti-localization\cite{Castro09}.

\begin{figure}[t]
\begin{center}
\includegraphics[width = 3in]{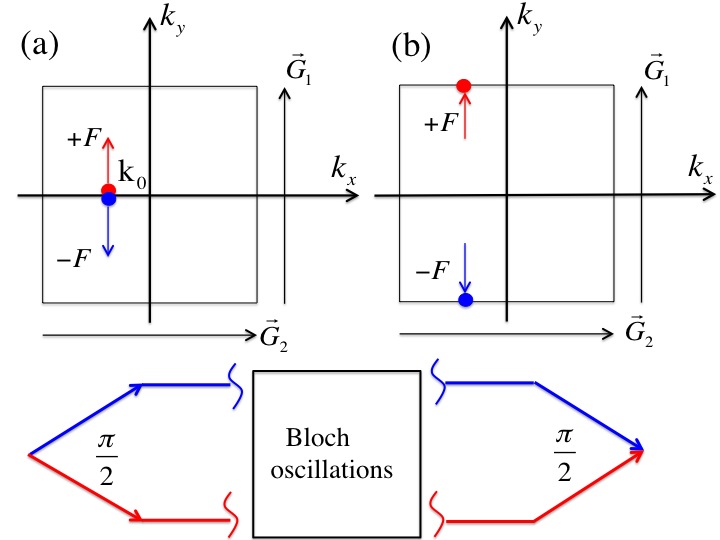}
\caption{Experimental setup for measuring the Zak phase. A cloud of ultracold atoms with a well-defined quasi-momentum ${\bf k_0}$ is loaded into a  2D optical lattice. ${\bf G_1}, {\bf G_2}$ denote reciprocal lattice vectors. Initially, atoms have spin up. (a) A $\pi/2$ pulse creates a coherent superposition of $|\!\! \uparrow \ra$ and $|\!\! \downarrow\ra$ states (marked by blue and red). After that, spin-selective forces $\pm {\bf F}$ parallel to ${\bf G_1}$ are applied. (b) After half a period of Bloch oscillations, when the two spins meet in the quasi-momentum space, another $\pi/2$ pulse is applied. The accumulated phase difference between the two states, which contains the Zak phase contribution, is measured by reading out the phase of the resulting Ramsey fringe.}
\label{figure0}
\end{center}
\end{figure}

There is currently a strong interest in realizing topological band structures in cold atomic systems~\cite{Bloch11,Soltan-Panahi2011,Tarruell2011,Becker,Spielman,Zwierlein}. Very recently, highly tunable 2D optical lattices with non-trivial Berry's phases have been demonstrated experimentally~\cite{Bloch11,Soltan-Panahi2011,Tarruell2011,Becker}. In particular, staggered flux lattice~\cite{Bloch11}, as well as brick-wall and honeycomb optical lattices with massless Dirac points have been engineered~\cite{Tarruell2011,Soltan-Panahi2011}. Moreover, several intriguing theoretical proposals of how to realize topologically non-trivial optical lattices with non-zero Chern number were put forward~\cite{Jaksch03,Mueller,Osterloh,Lim1,Gerbier,Lim2,Cooper11a,Cooper11b,Kolovsky,Kitagawa2010}. 

However, the traditional transport measurements, which can be used to determine the Chern number of a band via the quantized Hall effect, are very challenging in cold atomic systems. Thus, novel methods of probing the topology of bands in optical lattice are needed. One possible route to studying the topological structure of optical lattices relies on the fact that the Berry curvature gives rise to an anomalous velocity in semiclassical dynamics that can be monitored through in-situ images of the atom cloud~\cite{NiuReview,Price12}. Another possibility that has been pointed out recently is that time-of-flight images could be used to reveal topological invariants~\cite{Alba,Zhao}.

Here we propose an alternative, interferometric method for measuring Berry's phases, Berry curvature and Chern number of bands in 2D optical lattices. The idea is to combine Bloch oscillations with Ramsey interferometry for particles with two internal states, $|\!\!\uparrow\ra, |\!\!\downarrow\ra$, loaded into the optical lattice. The BZ has the topology of a torus, and therefore during the Bloch oscillations particles follow closed trajectories corresponding to the cycles of the BZ torus. The Berry's phases of such trajectories are known as the Zak phases~\cite{Zak89}.

We note that the two main ingredients required for the implementation of our proposal -- Bloch oscillations and Ramsey interferometry, have become standard techniques in cold atoms experiments~\cite{Salomon96,Inguscio04,Shin2004, Schumm2005,Gross2010, Widera2008}; thus, we expect that our proposal can be used to measure topological properties of 2D lattices in the near future. Very recently we have successfully applied a related approach to measure the Zak phases of topological Bloch bands in one dimension~\cite{1D}. 
 
 
We show how the measurements of the Zak phase allow one to obtain the more familiar topological characteristics of 2D Bloch bands. First, measuring the change of the Zak phase in the BZ, one can determine the distribution of the Berry curvature. This allows one to measure the Chern number of the band, given by the winding number of the Zak phase across BZ. Second, we show that the difference of Zak phases measured along certain trajectories can be linked to the Berry's phase of Dirac fermions. To be concrete, we consider the example of the brick-wall lattice~\cite{Tarruell2011}. However, all protocols that we propose can be extended to the cases of staggered flux lattice~\cite{Bloch11} and hexagonal lattice~\cite{Soltan-Panahi2011,Becker}.
 
 The scheme for measuring the Zak phase is the following (see Fig.~\ref{figure0}): Initially, a spin-up state with a given quasi-momentum ${\bf k_0}$ is prepared. The proposed protocol consists of three steps: (1) A $\pi/2$ pulse is used to create a coherent superposition of two spin  states, $ \left( |{\uparrow}\ra + |{\downarrow}\ra \right)/\sqrt{2}$; (2) Next, opposite forces $\pm {\bf F}$ on the spin-up and spin-down are applied, e.g., though a magnetic field gradient. It is required that the force ${\bf F}$ is parallel to some reciprocal lattice vector ${\bf G_1}$; (3) After half a period of the Bloch oscillations, when the two spins meet in quasi-momentum space, another $\pi/2$ pulse is applied and the $z$-component of the spin is measured. During such an evolution, the up and down states pick up a geometric contribution, equal to the Zak phase, which can be determined from the Ramsey phase.

{\bf Bloch oscillations and Zak phase.} We start our analysis by relating the Ramsey phase measured in the setup described above to the Zak phase. 
Consider a 2D lattice with the lattice vectors ${\bf a_1}, {\bf a_2}$, and the unit cell ${\bf r}\in x_1 {\bf a_1}+x_2 {\bf a_2}$, where $x_i \in [0,1)$. We denote the two primitive reciprocal lattice vectors by ${\bf G_1},{\bf G_2}$. According to the Bloch theorem, the eigenfunctions in the $n$th band can be represented in the following form
\be\label{eq:eigen_function}
\psi_{{\bf k}n}({\bf r})=e^{i {\bf k r}} u_{{\bf k}n}({\bf r}), 
\ee
where $u_{{\bf k} n}$ is the cell-periodic Bloch function, satisfying $u_{{\bf k} n} ({\bf r+a_i})=u_{{\bf k} n} ({\bf r})$, $i=1,2$. Notice that the $u_{{\bf k} n}$ is not periodic in the momentum space, but obeys the following condition: 
\be\label{eq:momentum_translate}
u_{{\bf k+G_i}n}({\bf r})= e^{-i {\bf G_ir}} u_{{\bf k}n}({\bf r}), 
\ee
which originates from the periodicity of the full Bloch function $\psi_{{\bf k}n}$.

\begin{figure}[t]
\begin{center}
\includegraphics[width = 2.5in]{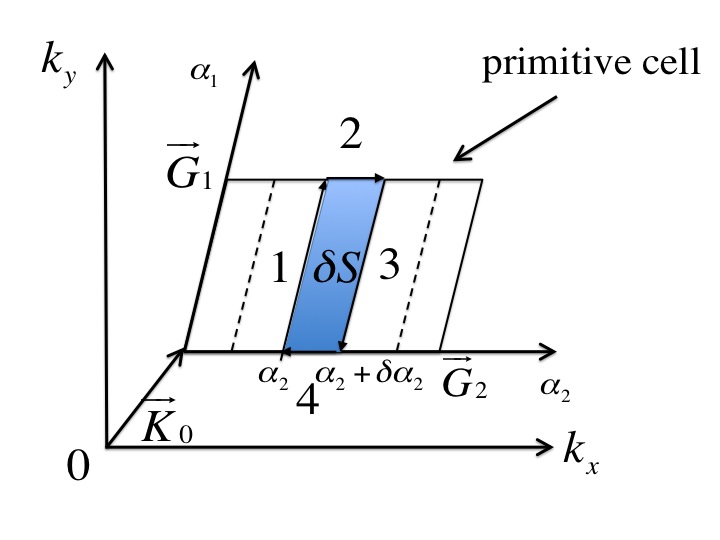}
\vspace{-0.7cm}
\caption{Primitive momentum-space cell of a generic 2D lattice. The Chern number of the band is related to the winding number of the Zak phase across the primitive cell, see Eq. (\ref{eq:chern}). The Zak phase can be measured using a combination of Bloch oscillations and Ramsey interferometry as described in the text.}
\label{figure4}
\end{center}
\end{figure}

We will assume that during the Bloch oscillations the particle motion remains adiabatic, such that the probability for the particle to get excited to another band is negligible. Following the first $\pi/2$ pulse, the particles are in the state $\psi_{{\bf k_0}n}({\bf r}) \otimes \frac{|\uparrow\ra+|\downarrow\ra}{\sqrt{2}}$. The evolution under the application of the force $\pm {\bf F}$ (opposite signs for opposite spins) is described by the time-dependent wave function $\Psi_{\uparrow} ({\bf r,t})\otimes \frac{| \uparrow\ra}{\sqrt{2}}+\Psi_{\downarrow} ({\bf r,t})\otimes \frac{|\downarrow\ra}{\sqrt{2}}$. The wave functions $\Psi_{\uparrow (\downarrow)} ({\bf r,t})$ obey the Schroedinger equation
\be\label{eq:schroedinger}
i\hbar\frac{\partial\Psi_{\uparrow} ({\bf r},t)}{\partial t}=H_{\uparrow(\downarrow)} \Psi_{\uparrow(\downarrow)}({\bf r,t}), 
\ee
with the Hamiltonian 
\be\label{eq:hamiltonian}
H_{\uparrow(\downarrow)}=H_0\mp {\bf F}{\bf r}\pm E_Z, \,\, H_0=-\frac{\hbar^2}{2m}\nabla^2+V({\bf r}), 
\ee
$V({\bf r})$ being the lattice potential, and $E_Z$ the Zeeman energy. 


As long as the adiabaticity condition is fulfilled, the particles remain within one band, and the solution of equation (\ref{eq:schroedinger}) has the following form:
\be\label{eq:psi}
\Psi_{\uparrow(\downarrow)} ({\bf r,t})=e^{i\xi _{\uparrow(\downarrow)}(t)} \psi_{{\bf k_{\pm}(t)},n} ({\bf r}), 
\ee
where ${\bf k_{\pm}}(t)={\bf k_0}\pm {\bf f}t$, ${\bf f}={\bf F}/\hbar$, and the phase $\xi_{\uparrow(\downarrow)}(t)$ is given by:
\be\label{eq:xi}
\xi_{\uparrow(\downarrow)}(t)=i\int_{\bf k_0} ^{{\bf k_{\pm}}(t)} \la u_{{\bf k'}n} | \nabla_{\bf k'} u_{{\bf k'}n}\ra d{\bf k'}-\frac{1}{\hbar} \int _0^t \epsilon_n ({\bf k_{\pm}}(t')) dt' \mp \frac{E_Z t}{\hbar}, 
\ee
The first term in the above equation describes the geometrical phase, while the second and third correspond to the dynamical phase, which depends on the speed of motion through the band. 

The Ramsey interferometry, performed after half a period of the Bloch oscillations (period is given by $T={G }/|{\bf f}|$), measures the phase difference picked up by the two spin species $\xi_{\uparrow}(T/2)-\xi_{\downarrow}(T/2)$. Using formula (\ref{eq:xi}), we obtain the Ramsey phase,
\be\label{eq:ramsey}
 \phi_{\rm tot}=\phi_{\rm Zak}+\phi_{\rm dyn}+\phi_{\rm Zeeman}, 
\ee
where the Zak phase is given by~\cite{Zak89}:
\be\label{eq:zak}
\phi_{\rm Zak}=i\int_{{\bf k_0-G}/2}^{{\bf k_0+G}/2} \la u_{{\bf k'}n} | \nabla_{\bf k'} u_{{\bf k'}n}\ra d{\bf k'}
\ee
and the dynamical phase and Zeeman phases are given by 
\be\label{eq:dynamical}
\phi_{\rm dyn}=- \frac{1}{\hbar} \int _{-T/2}^{T/2} {\rm sign}(t') \epsilon_n ({\bf k_0}+{\bf f} t') dt', \, \phi_{\rm Zeeman}=-\frac{E_ZT}{\hbar}. 
\ee
For the case of a band structure with symmetric dispersion relation, $\epsilon_n ({\bf k_0+{\bf f}}t')=\epsilon_n ({\bf k_0-{\bf f}}t')$, the dynamical phase vanishes, and the Ramsey interferometry directly gives the Zak phase. This is the case for special choices of ${\bf k_0}$ and ${\bf G_1}$ in the experimentally relevant case of the brick-wall lattice which we will discuss below. 

{\bf Measuring Berry curvature and Chern number of a generic band.} Let us now turn to the discussion of how Ramsey interferometry can be used to determine the Berry curvature and the Chern number (and therefore the topological class) of a gapped band; no special symmetries are assumed, except for the symmetry of dispersion which guarantees the cancellation of the dynamical phase, and allows the separation of the Zak phase. 

We choose the primitive cell in quasi-momentum space to be a torus defined by ${\bf k}={\bf K_0}+\alpha_1 {\bf G_1}+\alpha_2 {\bf G_2}$, where $\alpha_i\in [0;1)$ and ${\bf K_0}$ is an arbitrary quasi-momentum (as shown in Fig.~\ref{figure4}). We notice that the Chern number cannot be determined by measuring the Zak phases along the four sides of the torus, essentially, because the Zak phase is only defined modulo $2\pi$. However, as we now discuss, the Chern number ${\cal{C}}$ can be related to the {\it winding} number of the Zak phase across the BZ (see Ref.~\cite{NiuReview} for a closely related discussion in the context of adiabatic pumping). 

\begin{figure}[t]
\begin{center}
\includegraphics[width = 3.2in]{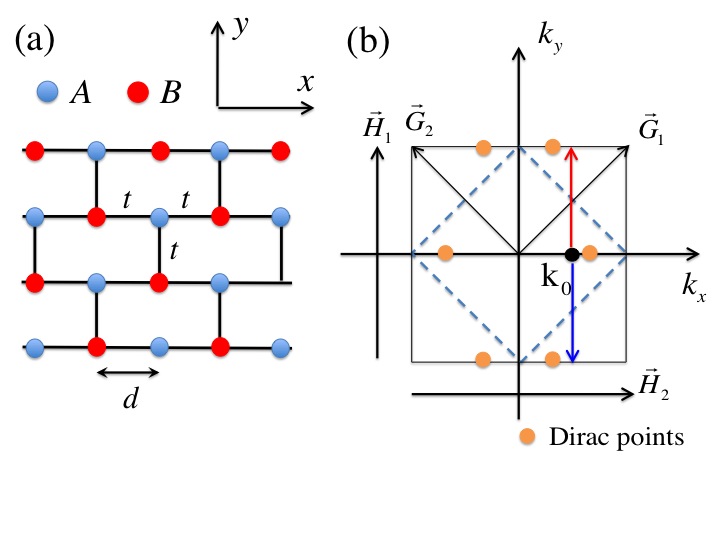}
\vspace{-0.7cm}
\caption{a) Brick-wall lattice. $A$ and $B$ sites are marked by blue and red circles, nearest-neighbor hopping is assumed.  
b) The Brillouin zone of the brick-wall lattice model (blue dashed square). The band structure exhibits two Dirac points marked by orange circles. Owing to the symmetry of the dispersion, it is convenient to measure the Zak phase with initial quasi-momentum ${\bf k_0}=(k_0,0)$ lying on the $x$ axis, and applying a force in the $y$ direction. Measuring the variation of the Zak phase as a function of $k_0$, it is possible to (i) measure the $\pi$ Berry's phase of Dirac particles, (ii) measure the  Chern number of the bands when they are separated by energy gaps.}
\label{figure1}
\end{center}
\end{figure}

We consider an experiment in which the Zak phase is measured for torus cycles defined by ${\bf G_1}$ as a function of $\alpha_2$, see Fig.~\ref{figure4}. Experimentally, this would be achieved by preparing the initial state ${\bf k_0}={\bf K_0}+{\bf G_1}/2+\alpha_2 {\bf G_2}$ for different values of $\alpha_2$. 

Let us show that the small change of Zak phase as $\alpha_2$ is increased by $\delta \alpha_2$ is equal to the integral of the Berry curvature over the rectangle $\delta S$ defined by the corresponding trajectories (see Fig.~\ref{figure4}). Equivalently, the difference of the Zak phases $\gamma=\phi_{\rm Zak}(\alpha_2+\delta\alpha_2)-\phi_{\rm Zak}(\alpha_2)$ is given by the Berry's phase that corresponds to the contour 1234. It is easiest to see this by choosing a smooth gauge for the periodic Bloch function in $\delta S$ (this can be done since region $\delta S$ is small; in general, no smooth gauge can be chosen in the whole BZ). The Berry's phase $\gamma$  can be represented as the sum of the Berry's phases for the four sides of the rectangle, $\gamma=\sum_{i=1}^4 \gamma_i$. Since the sides $2$ and $4$ are equivalent (they differ by ${\bf G_1}$), but are traversed in the opposite direction, their contribution vanishes, $\gamma_2+\gamma_4=0$. Because we chose the periodic gauge, $\gamma_3+\gamma_1$ is equal to the difference of the Zak phases for trajectories $3$ and $1$. Thus, the change of the Zak phase is related to the Berry's phase, which can be written as an integral of the Berry's curvature $\Omega_{12}$, $\gamma=\int_{\delta {\cal S}} d^2 k  \Omega_{12}({\bf k}).$ 
This relation can be conveniently written in terms of a uniquely defined quantity $z(\alpha_2)=e^{i\phi_{\rm Zak} (\alpha_2)}$: 
\be\label{eq:gamma2}
\int_{\delta {\cal S}} d^2 k  \Omega_{12}({\bf k})=-i{z}^*(\alpha_2) \partial_{\alpha_2} z(\alpha_2) \delta \alpha_2. 
\ee 
Summing relation (\ref{eq:gamma2}) over different regions, and using the definition of the Chern number ${\cal{C}}=\frac{1}{2\pi} \int_{\rm BZ} d^2 k  \Omega_{12}({\bf k})$, we then obtain $c$ via the winding of the Zak phase,  
\be\label{eq:chern}
{\cal{C}}=-\frac{i}{2\pi} \int _0^1 d\alpha_2 {z}^*(\alpha_2) \partial_{\alpha_2}{z}(\alpha_2).
\ee
 This relation implies that the interferometric measurements of the Zak phase across the BZ allow the extraction of the Chern number, and can be used to detect the topological nature of the bands and topological phase transitions.  

{\bf Brick-wall lattice: measuring $\pi$ Berry's phase and detecting topological phase transitions. } Our approach is also suitable for measuring the $\pi$ Berry's phase of the massless Dirac fermions in non-trivial lattices~\cite{Bloch11,Soltan-Panahi2011,Tarruell2011}. For definiteness, we consider the brick-wall lattice~\cite{Tarruell2011}, illustrated in Fig.~\ref{figure1}. 

The particles in the brick-wall lattice with nearest-neighbor hoppings $t$ are described by the Hamiltonian $H=-t\sum_{\la ij \ra} c_i^{\dagger} c_j$ (the analysis below can be straightforwardly generalized to the case of anisotropic hopping in the $x$ and $y$ direction). Assume that an $A$-site coincides with the origin (this choice is important, because it fixes the periodicity condition (\ref{eq:momentum_translate}), see Fig.~\ref{figure1}).  The cell-periodic wave function, given by a two-component vector $(v_{A{\bf k}}, v_{B{\bf k}})$ (corresponding to the amplitudes on the sub-lattices $A$, $B$), can be found from the Schroedinger equation with an effective Hamiltonian
\be\label{eq:hamiltonian_graphene} 
H_{\bf {k}} =  -   
\left( \begin{array}{cc} 0 &t_{\bf k} \\ t^{*}_{\bf {k}} & 0 \end{array} \right)  ,  
\ee
\be\label{eq:t}
t_{\bf k}  = t (2 \cos k_x d +e^{-i k_y d} ) =\epsilon_{\bf k} e^{i\theta_{\bf k}},
\ee
where we have chosen the vector connecting $A$-site to its neighbors to be equal to $1$, and the energy of the Bloch states is given by $\epsilon=\pm \epsilon_{\bf k}$, 
\be\label{eq:energy}
\epsilon_{\bf k}=t \sqrt{4 \cos^2 k_xd +4 \cos k_xd \cos k_yd +1}.
\ee
The eigenfunctions for the valence and conduction bands are given by $(v_{A{\bf k}},v_{B{\bf k}})=\frac{1}{\sqrt{2}}(1, \pm e^{-i\theta_{\bf k}}).$ In principle, the wave functions are defined up to a gauge transformation; here the gauge is fixed such that the wave functions satisfy the periodicity condition (\ref{eq:momentum_translate}). 

The BZ, illustrated in Fig.~\ref{figure1}, is defined by primitive reciprocal lattice vectors ${\bf G_1}=(\pi, \pi)$ and ${\bf G_2}=(-\pi,\pi)$. However, measuring the Zak phases for the cycles defined by ${\bf G_{1,2}}$ is complicated by the lack of corrsponding symmetry of the energy dispersion; thus, it would not be feasible to cancel the dynamical phase. 

To overcome this difficulty, we propose to measure Zak phases for ${\bf k_0}$ lying on the $x$ axis, with the force applied in the $y$ direction. This allows for a measurement of the Zak phases corresponding to the reciprocal lattice vector ${\bf H_1}={\bf G_1}+{\bf G_2}$ (see Fig.~\ref{figure1}), while cancelling the dynamical phase (as guaranteed by the $k_y\to -k_y$ symmetry of the dispersion (\ref{eq:energy})). Similarly, it is possible to measure the Zak phases for trajectories corresponding to reciprocal vector ${\bf H_2}={\bf G_1}-{\bf G_2}$, when the initial momentum ${\bf k_0}$ lies on the $y$ axis. In this case, the dynamical phase cancels out due to $k_x \to -k_x$ symmetry of the dispersion.

Measuring the Zak phases across the BZ in this case allows direct detection of the $\pi$ Berry's phase of the Dirac fermions. As the initial momentum ${\bf k_0}$ passes either of the two Dirac points, situated at $(\pm 2\pi/3,0)$, the Zak phase must jump by $\pi$.  

Furthermore, the method described above can be used to determine the topological nature of gapped bands in the brick-wall lattice. The winding number of the Zak phase measured as a function of initial momentum ${\bf k_0}$ on the $x$-axis gives {\it twice} the Chern number (repeating the argument above, one can show that Berry's phase winding is equal to the integral of Berry's curvature over a region $-\pi < k_x.k_y <\pi$, which contains two Brillouin zones). We note that both topologically trivial and non-trivial gaps can be introduced by adding extra terms in the Hamiltonian (\ref{eq:hamiltonian_graphene}). The topologically trivial gaps arise, e.g., when a staggered potential between $A$ and $B$ sites is turned on. In this case, the Chern number is zero. The Haldane model, in which next-nearest-neighbor hoppings with non-trivial phases $\pm \eta$ are turned on~\cite{Haldane88}, provides a realization of a non-trivial band, with the Chern number $\pm 1$. Importantly, both models still preserve the symmetry of dispersion with respect to $k_x \to -k_x$, $k_y \to -k_y$ (the Haldane model obeys this condition when $\eta =\pm \pi/2$). Thus, the problem of dynamical phases does not occur when the Zak phases (and Chern numbers) are measured. 

{\bf Protocols insensitive to fluctuating magnetic fields.} Now we would like to address an important experimental concern related to the presence of slowly fluctuating magnetic fields. The fluctuations of the magnetic field induce shot-to-shot variations in $E_Z$ and $\phi_{\rm Zeeman}$, complicating the reliable extraction of the Zak phase. We propose various protocols which are insensitive to the magnetic noise. Here we briefly mention the main ideas, and provide a detailed description in Ref.~\cite{SOM}. 

The first protocol allows measurements of the {\it variations} of the Zak phase across the BZ (and in fact, only the difference of the Zak phase is physically meaningful, while the Zak phase itself depends on the choice of the real-space unit cell, see, e.g.~\cite{1D}). The idea is to prepare the system in two or more different quasi-momentum states, and to carry out the sequence described above for each initial state. Assuming that the dispersion is symmetric, the difference of the phases extracted from Ramsey fringes will be equal to the difference of the Zak phases. This protocol therefore allows one to obtain the Berry curvature and the Chern number of the band. It is insensitive to the shot-to-shot fluctuations of the magnetic field because the Zeeman phases picked up by different quasi-momentum states are equal. A more detailed discussion of this protocol for the brick-wall lattice can be found in Ref.\cite{SOM}. Another approach is to design protocols that combine Bloch oscillations with {\it spin echo} sequence~\cite{SOM}. Such protocols are naturally insensitive to the fluctuating magnetic fields.



{\bf Summary}. In conclusion, we have presented an approach to studying topological properties of 2D optical lattices. This approach allows one to measure the Berry's phase of the Dirac fermions, as well as the local Berry curvature and the Chern number of the band. The latter measurement can be used to study the topological structure of the bands in optical lattices and to detect the topological phase transitions that occur as a function of the lattice parameters. 


{\bf Acknowledgements.} We thank T. Esslinger for insightful discussions. The authors acknowledge support from a grant from the Army Research Office with funding from the DARPA OLE program, Harvard-MIT CUA, NSF Grant No. DMR-07-05472, AFOSR Quantum Simulation MURI, the ARO-MURI on Atomtronics.

%

\onecolumngrid
\section{Supplementary online material for "Interferometric approach to measuring band topology in 2D optical lattices"}






\begin{figure}[t]
\begin{center}
\includegraphics[width = 3.4in]{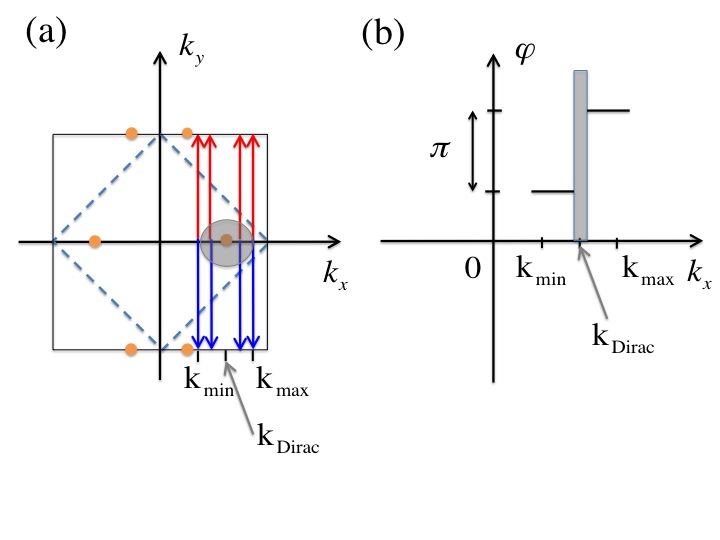}
\caption{A protocol in which a variation of Zak phase across BZ in measured, for the case of a brick-wall lattice. (a) An ensemble of atoms with different quasi-momenta is loaded into an optical lattice (grey cloud). The Ramsey/Bloch sequence allows measurement of the Zak phase for quasi-momenta in the interval $k_{\rm min}<k_x<k_{\rm max}$. (b) The Zak phase exhibits a $\pi$ jump when $k_x$ crosses the Dirac point (situated at $k_x=2\pi/3$), which provides a signature of the $\pi$ Berry's phase of Dirac fermions.}
\label{figure5}
\end{center}
\end{figure}

Here, we provide a detailed description of the protocols which allow one to eliminate the effect of fluctuating magnetic fields. We illustrate the ideas by using the experimentally relevant example of brick-wall lattice, noting that extensions to the case of flux lattice~\cite{Bloch11} and honeycomb lattice~\cite{Soltan-Panahi2011} are straightforward. 

\begin{figure}[t]
\begin{center}
\includegraphics[width = 3.4in]{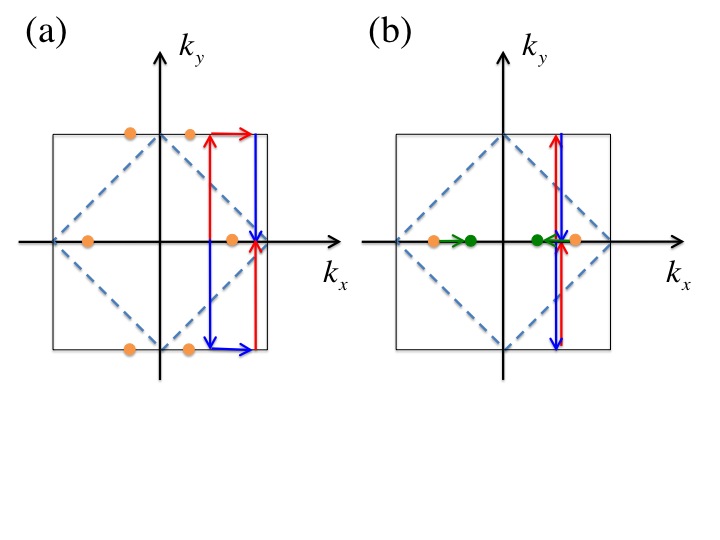}
\vspace{-0.5cm}
\caption{(a) Spin-echo protocol for measuring the Berry's phase of the Dirac fermions and Chern number of the band. (b) Modified spin-echo protocol for measuring the Berry's phase of the Dirac fermions in brick-wall lattice.}
\label{figure6}
\end{center}
\end{figure}

The first protocol (which was discussed in the main text in some detail) is illustrated in Fig.~\ref{figure5}. One can envision that this protocol can be implemented using either bosons or fermions. For the case of bosons, one would prepare two clouds of atoms with different quasi-momenta $k_x =k_{\rm min}$, $k_x= k_{\rm max}$ and $k_y=0$. We cannot predict the actual Ramsey/Bloch phases in a single run of the experiment due to magnetic field fluctuations which lead to fluctuations of the Zeeman contribution. However, the difference of the phases should be $\pi$ when the two quasi-momentum states lie on the opposite sides of the Dirac point, as in Fig.~\ref{figure5}. Note that particle crossing the Dirac point do not have a well defined Ramsey/Bloch phase due to the breakdown of adiabaticity condition.
Alternatively, an ensemble of cold fermionic atoms occupying a finite region in quasi-momentum space, is loaded into a brick-wall lattice. The Ramsey/Bloch phases can be measured for each momentum ${\bf k}$ separately. Thus we can compare accumulated Ramsey/Bloch phases for different initial momenta. This allows measurement of the $\pi$ Berry's phase of Dirac fermions, and, more generally, the Berry's curvature. 


Another approach is to combine Bloch oscillations with spin-echo sequence. Such protocols are naturally insensitive to fluctuating magnetic fields, because each atom spends equal time being in the spin-up and spin-down state. Therefore, the Zeeman phases picked up by all atoms are equal and cancel each other out.

One possible spin-echo-type protocol, illustrated in Fig.~\ref{figure6}(a), consists of three steps: (1) A $\pi/2$ pulse, combined with half a period of Bloch oscillations, similar to the sequence for measuring the Zak phase discussed in the main text. (2) Then, both spin components are quickly moved in the $y$ direction (applied forces are equal). This is followed by the $\pi$ pulse, which flips spins up and down. (3) Another half a period of Bloch oscillations is performed. (Notice that the direction of the forces applied to pseudo-spin up and down is the same as in step (1).) After that, a $\pi/2$ pulse is applied and the accumulated phase is measured. When the trajectory of the particle encloses the Dirac point (as shown in Fig.~\ref{figure6}(a)), the accumulated phase is equal to $\pi$ (the Berry's phase of Dirac fermions). We emphasize that this protocol is not limited to the case of brick-wall lattice, and in general allows the measurement of the band's Berry's curvature and the Chern number. 

Another possible protocol (which we call modified spin-echo) uses the experimental ability to tune the band structure {\it in situ}~\cite{Tarruell2011}. The idea of the protocol, illustrated in Fig.~\ref{figure6}(b), is not to move atoms in the same direction (step (2) described above), but rather to move Dirac points by dynamically changing the band structure. The protocol consists of three steps: (1) A $\pi/2$ pulse followed by half a period of Bloch oscillations. (2) A $\pi$ pulse. (3) Half a period of Bloch oscillations, followed by $\pi/2$ pulse and measurement of the Ramsey phase. During steps (1),(2),and (3) parameters of the optical lattice are changed adiabatically such that atoms return on the other side of the Dirac point. This measurement gives the $\pi$ Berry's phase of the Dirac fermions.


\begin{thebibliography}{99}

\bibitem{TKNN}
D. J. Thouless, M. Kohmoto, M. P. Nightingale, and M. den Nijs, Phys. Rev. Lett. 49, 405 (1982). 

\bibitem{NiuReview}
D. Xiao, M. C. Chang, and Q. Niu, Rev. Mod. Phys. 82, 1959 (2010). 

\bibitem{Niu95}
M.-C. Chang and Q. Niu, Phys. Rev. Lett. 75, 1348 (1995).

\bibitem{Haldane2004}
F. D. M. Haldane, Phys. Rev. Lett. 93, 206602 (2004). 

\bibitem{Berry}
M. V. Berry, Proc. Roy. Soc. London A 392, 451 (1984).

\bibitem{Castro09}
A. H. Catro Neto {\it et al.}, Rev. Mod. Phys. 81, 109 (2009). 

\bibitem{Hasan}
M. Z. Hasan, C. L. Kane, Rev. Mod. Phys. 82, 3045 (2010). 

\bibitem{Bloch11}
M. Aidelsburger {\it et al.}, Phys. Rev. Lett. 107, 255301 (2011). 

\bibitem{Soltan-Panahi2011} 
P. Soltan-Panahi {\it et al.}, Nature Physics 7, 434 (2011). 

\bibitem{Tarruell2011}
L. Tarruell {\it et al.}, Nature 483, 302 (2012). 

\bibitem{Becker} C. Becker, P. Soltan-Panahi, J. Kronj\"ager, S. D\"orscher, K. Bongs, K. Sengstock, New J. Phys. 12, 065025 (2010).

\bibitem{Spielman} Y.-J. Lin, K. Jimenez-Garcia, I. B. Spielman, Nature 471, 83 (2011). 

\bibitem{Zwierlein} Lawrence W. Cheuk, Ariel T. Sommer, Zoran Hadzibabic, Tarik Yefsah, Waseem S. Bakr, Martin W. Zwierlein, 
arXiv:1205.3483 (2012). 

\bibitem{Jaksch03} D. Jaksch, P. Zoller, New Journal of Physics 5, 56 (2003). 

\bibitem{Mueller} E. Mueller, Phys. Rev. A 70, 041603 (2004).

\bibitem{Osterloh} K. Osterloh, M. Baig, L. Santos, P. Zoller, M. Lewenstein, 
Phys. Rev. Lett. 95, 010403 (2005).

\bibitem{Lim1} L.-K. Lim, C. Morais Smith, Andreas Hemmerich, Phys. Rev. Lett. 100 130402 (2008).

\bibitem{Gerbier} F. Gerbier, J. Dalibard, New J. Phys. 12, 033007 (2010). 

\bibitem{Lim2} L.-K. Lim, A. Lazarides, A. Hemmerich, and C. Morais
Smith, Phys. Rev. A 82 013616 (2010). 

\bibitem{Cooper11a} N. R. Cooper, Phys. Rev. Lett. 106, 175301 (2011).

\bibitem{Cooper11b} N. R. Cooper, J. Dalibard, Europhys. Lett. 95, 66004 (2011).

\bibitem{Kolovsky} A. Kolovsky, Europhys. Lett. 93, 20 003 (2011). 

\bibitem{Kitagawa2010}
T. Kitagawa, E. Berg, M. Rudner, E. Demler, Phys. Rev. B 82, 235114 (2010). 

\bibitem{Price12}
H. M. Price, N. R. Cooper, Phys. Rev. A 85, 033620 (2012). 

\bibitem{Alba}  E. Alba et al., Phys. Rev. Lett. 107, 235301 (2011).

\bibitem{Zhao}  E. Zhao et al., Phys. Rev. A 84, 064629 (2011).

\bibitem{Zak89}
J. Zak, Phys. Rev. Lett. 62, 2747 (1989). 

\bibitem{Salomon96}
M. B. Dahan, E. Peik, J. Reichel, Y. Castin, and C. Salomon, 
Phys. Rev. Lett. 76, 4508 (1996). 

\bibitem{Inguscio04}
G. Roati, E. de Mirandes, F. Ferlaino, H. Ott, G. Modugno, and M. Inguscio, 
Phys. Rev. Lett. 92, 230402 (2004). 

\bibitem{Shin2004}
Y. Shin {\it et al.}, Phys. Rev. Lett. 92, 050405 (2004). 

\bibitem{Schumm2005}
T. Schumm {\it et al.}, Nature Physics 1, 57 (2005). 

\bibitem{Gross2010}
C. Gross {\it et al.}, Nature 464, 1165 (2010). 

\bibitem{Widera2008}
A. Widera {\it et al.}, Phys. Rev. Lett. 100, 140401 (2008). 

\bibitem{1D}
M. Atala {\it et al.}, submitted. 

\bibitem{Haldane88}
F. D. M. Haldane, Phys. Rev. Lett. 61, 2015 (1988). 


\bibitem{SOM}
Supplementary Online Material. 



\end{thebibliography}
\end{document}